\documentclass[9pt,twocolumn,twoside]{opticajnl}
\journal{opticajournal} 

\setboolean{shortarticle}{false}

\usepackage{color}
\usepackage{ulem}
\usepackage{bm}  

\newcommand {\scat} {\mathrm{sc}}

\title{Light propagation in atomic stratified media: breakdown of the transfer-matrix method at high density}

\author[1]{Igor M. Sokolov}
\author[2]{William Guerin}

\affil[1]{Ioffe Institute, 194021, St.-Petersburg, Russia}
\affil[2]{Universit\'e C\^ote d'Azur, CNRS, Institut de Physique de Nice, France}

\begin{abstract}
The transfer-matrix method is a standard approach to wave propagation in stratified media. With the advent of cold-atom-based quantum and photonic technologies, several experiments and many proposals consider light propagation in one-dimensional optical lattices, using the transfer matrices as the main tool for the simulation. Here, we study the validity of this method by comparing its results to the microscopic coupled-dipole model, which is exact in the linear-optics regime. We show that the transfer-matrix method works very well at low density, even for thin disordered slices, and breaks down at high density because the dipole-dipole interaction induces a collective response from the atoms such that the properties of one layer are influenced by the others. We determine the boundary values of atomic densities for which this method is still applicable for describing experiments. Our findings are relevant for experimental realizations using ultra-cold atoms. 
\end{abstract}

\setboolean{displaycopyright}{false} 

\begin{document}

\maketitle

The collective response of cold atomic samples interacting with light is a topic of current active research, for fundamental reasons and for applications \cite{Guerin:2017a, Chang:2018, Reitz:2022, Sheremet:2023, Jen:2025}. Among the fundamental problems, the understanding of the value of the refractive index of gases \cite{Andreoli:2021} and of their resonance line shape and widths has been the subject of intense debate (see \cite{Fofanov:2011, Javanainen:2016, Jennewein:2018, Vatre:2024} and references therein). The applications are in the field of quantum technologies, for instance clocks \cite{Chang:2004}, quantum memories \cite{Asenjo:2017}, new lasing schemes \cite{Guerin:2008, Bohnet:2012, Schilke:2012a, Baudouin:2013a, Holzinger:2020}, etc.

When microscopic samples, made of a small number of atoms, are considered, the modeling can rely on exact numerical solutions or, in the linear-optics regime, on the coupled-dipole equations \cite{Javanainen:1999, Svidzinsky:2008, Sokolov:2011}. This becomes impracticable for very large samples containing millions of atoms or more, and in many cases it is very useful to have macroscopic theories. For instance, the propagation through a disordered sample can be computed from Beer-Lambert exponential law. Although extremely simple, it is enough to explain some observed collective effects on the amount of scattered light \cite{Pellegrino:2014,Sokolov:2019,Kemp:2020}, on the radiation pressure force \cite{Bachelard:2016}, or on collective optical transients \cite{Chalony:2011,Kwong:2014,Kwong:2015}. For optically dense samples, multiple scattering of light is an essential ingredient, and it can be simulated by a random-walk approach, which proved good enough to explain a lot of observables, in the stationary regime \cite{Kupriyanov:2003, Labeyrie:2004} and even for the temporal dynamics \cite{Labeyrie:2003, Datsyuk:2006}. Remarkably, a mixture of Beer-Lambert propagation in an effective medium and multiple scattering could also explain superradiance and subradiance in the linear-optics regime \cite{Guerin:2016a, Araujo:2016, Kuraptsev:2017, Weiss:2021, Asselie:2022, Guerin:2023}. Finally, for stratified samples, for instance cold atoms trapped in a one-dimensional (1D) lattice, a transfer-matrix model can be used \cite{Deutsch:1995, Slama:2006, Petrosyan:2007, Schilke:2011, Asselie:2025}.

These approaches are very fruitful at low density $\rho \ll k^3$, when one can use textbook formulas as building blocks of the macroscopic theories. For example, the scattering cross-section of individual atoms, $\sigma_\scat = (k^4/6\pi)|\alpha|^2$, where $\lambda=2\pi/k$ is the wavelength of the transition, is the essential ingredient of the random-walk model for multiple scattering. Here $\alpha$ is the atomic polarizability, 
\begin{equation}\label{eq.alpha}
\alpha = \frac{6\pi}{k^3} \times \frac{2\Delta/\Gamma + i}{1+4 \Delta^2/\Gamma^2} \,,
\end{equation}
which also allows computing the complex refractive index,
\begin{equation}\label{eq.n}
n = \sqrt{1 + \rho \alpha} \,,
\end{equation}
which is the building block of Beer-Lambert law and the transfer-matrix model.

However, at larger densities, when the typical distance between atoms is smaller than the wavelength, these simple formulas break down because of the strong near-field dipole-dipole interaction (DDI) between atoms. This has been known for a long-time, with for instance the textbook's Lorentz-Lorenz shift \cite{BornWolf}. However, recent research have shown that the textbook result does not apply to cold atoms because of the strong correlations induced by the DDI \cite{Fofanov:2011,Javanainen:2016,Andreoli:2021}.


In this article, we study the validity range of the transfer-matrix (TM) model to simulate the propagation and reflection of light, in the linear-optics regime, through a 1D lattice of cold atoms. By comparing with the coupled-dipole (CD) model, we show that the TM model  using textbook formulas for the refractive index works well at low density, even for a very thin (subwavelength) slice, and, not surprisingly, breaks down at large density. However, we also show that using the effective refractive index determined at high density from the CD model as the ingredient of the TM model does not allow to restore its validity. We interpret this less intuitive result as being due to the DDI interaction between layers.

The article is organized as follows. We first present the transfer matrix model and the coupled-dipole model in section \ref{sec.models}. We then compare the results of the two models in section \ref{sec.comparison}, first in the low density regime, then at higher density for an individual thick layer. This allows us to identify the domain of validity of the TM model. Finally in section \ref{sec.Bragg} we apply our findings to the specific and interesting case of a Bragg mirror made of cold atoms.

\section{Description of the models}\label{sec.models}

\subsection{Transfer-matrix method}\label{sec.TM}

\begin{figure}[b]
\centering
\includegraphics{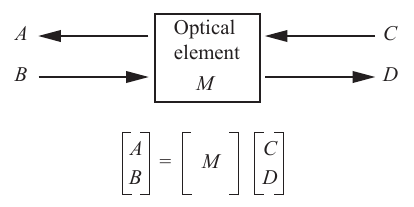}
\caption{Illustration of the transfer-matrix method. $A,B,C,D$ are the amplitudes of the fields.}
\label{fig.TM}
\end{figure}

The TM method relates forward- and backward-traveling waves on the right-hand side of an arbitrary optical element to those on the left-hand side \cite{BornWolf} (Fig.\,\ref{fig.TM}). It has been used for the first time in the context of cold atomic lattices by Deutsch \textit{et al.} \cite{Deutsch:1995}. It is based on the Fresnel coefficients at each interface and on the Beer-Lambert attenuation law for the transmission inside the atomic layer. These two ingredients are computed from the complex refractive index (\ref{eq.n}). At normal incidence, the Fresnel coefficients of the vacuum to atomic layer interface are 
\begin{equation}
r_\mathrm{va} = \frac{1-n}{1+n}  \quad \& \quad t_\mathrm{va} = \frac{2}{1+n} \,,
\end{equation}
and for the interface from the atomic layer to vacuum:
\begin{equation}
r_\mathrm{av} = \frac{n-1}{n+1}  \quad \& \quad t_\mathrm{av} = \frac{2n}{1+n} \,.
\end{equation}
Finally, the transmission through the atomic layer of thickness $L$ is
\begin{equation}
t_\mathrm{a} = e^{inkL} \,.
\end{equation}

From the definition of the transfer matrix as given by Fig.\,\ref{fig.TM}, the matrix of each interface is \cite{Klein:1986} ($i=\{\mathrm{av}, \mathrm{va}\}$):
\begin{equation}\label{eq.M}
M_i = \frac{1}{t_i} \begin{bmatrix} 1 & r_i \\ r_i & 1 \end{bmatrix} \,,
\end{equation}
and the transfer matrix through the atomic layer is:
\begin{equation}\label{eq.Ma}
M_\mathrm{a} = \begin{bmatrix} t_\mathrm{a} & 0 \\ 0 & 1/t_\mathrm{a} \end{bmatrix} \,.
\end{equation}
Obviously the same matrix can be used for a vacuum gap between layer by replacing $n$ by 1.

Therefore, the total matrix of a layer of atoms surrounded by vacuum is
\begin{equation}
M = M_\mathrm{va} M_\mathrm{a} M_\mathrm{av} \,,
\end{equation}
and the global transmission and reflection coefficient for the field are
\begin{equation}\label{eq.t_M}
r_M = \frac{M_{12}}{M_{22}} \, , \quad t_M = \frac{1}{M_{22}} \,.
\end{equation}
A little bit of algebra gives
\begin{equation}\label{eq.t}
t(n) = \frac{ t_\mathrm{va} t_\mathrm{av} t_\mathrm{a} }{1+ r_\mathrm{va} r_\mathrm{av} t_\mathrm{a}^2 } \, ,
\end{equation}
\begin{equation}\label{eq.r}
r(n) = \frac{ r_\mathrm{va} + r_\mathrm{av} t_\mathrm{a}^2 }{1+ r_\mathrm{va} r_\mathrm{av} t_\mathrm{a}^2 } \, .
\end{equation}
Here we have explicitly reminded the dependence with the complex refractive index $n$.

This method is particularly powerful for stratified media, like thin films. If the medium is periodic, like for a Bragg mirror (sec.\,\ref{sec.Bragg}), the whole transfer matrix is a power of the matrix of a single period.

\subsection{Coupled-dipole equations}\label{sec.CD}


The CD model, widely used in the last years in the context of collective effects in light-atom interactions \cite{Guerin:2017a, Chang:2018, Reitz:2022, Sheremet:2023, Jen:2025}, considers $N$ two-level atoms (ground state with the total angular momentum $J_g = 0$, and degenerate excited state $J_e = 1$ with $m\equiv J_z = -1,0,1$) at random positions $\bm{r}_i$ driven by an incident laser (Rabi frequency $\Omega(\bm{r})$, detuning $\Delta$).
Restricting the Hilbert space to the subspace spanned by the ground state of the atoms $|G\rangle = |g \cdots g \rangle$ and the singly-excited states $|i\rangle = |g \cdots e_{im} \cdots g\rangle$ and tracing over the photon degrees of freedom, one obtains an effective Hamiltonian describing the time evolution of the atomic wave function $| \psi(t) \rangle$,
\begin{equation}
| \psi(t) \rangle = \alpha(t) | G \rangle +  \sum\limits_{i=1}^N \sum\limits_{m} \beta_{e_{im}}(t)| i \rangle \; . \label{eq:psi}
\end{equation}
Considering the low intensity limit, when atoms are mainly in their ground states, i.e. $\alpha \simeq 1$, the problem amounts to determine the amplitudes $\beta_{e_{im}}$, which are then given by the linear system of coupled equations
\begin{equation}
\dot{\beta}_{e_{im}} = \left( i\Delta-\frac{\Gamma}{2} \right)\beta_{e_{im}} -\frac{i\Omega_{e_{im}}}{2} + \frac{i\Gamma}{2} \sum_{j \neq i}\sum\limits_{m'} V_{e_{im}e_{jm'}}\beta_{e_{jm'}}\; .
\label{eq.betas}
\end{equation}
These equations are the same as those describing $N$ classical dipoles driven by an oscillating electric field~\cite{Foldy:1945, Lax:1951, Svidzinsky:2010}.
The first term on the right hand side corresponds to the natural evolution of independent dipoles (with $\Gamma$ the linewidth of the transition), the second one to the driving by the external laser, the last term corresponds to the dipole-dipole interaction and is responsible for all collective effects. It reads
\begin{eqnarray}\label{eq.ddi}
V_{e_{im}e_{jm'}}& =&
-\frac{4\pi k^2}{\hbar}  \sum_{\mu\nu}\mathbf{d}_{e_{im} g_i}^{\mu} \mathbf{d}_{g_j e_{m'j}}^{\nu}
G_{\mu\nu}(\mathbf{r}_{ij}) \;.
\end{eqnarray}
Here $\mathbf{d}_{e_{im} g_i}$ is the dipole moment operator of the transition ${g} \to {e_m}$ of the atom $i$, $\mathbf{r}_{ij} =\mathbf{r}_i - \mathbf{r}_j$ and $r_{ij} = |\bm{r}_i - \bm{r}_j|$. The superscripts $\mu$ or $\nu$ denote projections of vectors on one of the reference frames,
\begin{eqnarray}\label{eq.green}
G_{\mu\nu}(\mathbf{r}) = -\frac{e^{i k r}}{4 \pi r}
\left[ P(i k r) \delta_{\mu\nu}
+ Q(ik r) \frac{\mathbf{r}_\mu  \mathbf{r}_\nu}{r^2} \right]
\end{eqnarray}
is the  Green's matrix of Maxwell equations with $k = \omega_0/c = 2\pi/\lambda_0$, $P(u) = 1 - 1/u + 1/u^2$, $Q(u) = -1 + 3/u - 3/u^2$.

To calculate the reflectivity and refraction coefficients of an infinitely wide slab, we use the approach described in \cite{Sokolov:2015, Skipetrov:2025}. We consider cylindrical samples of radius $R \gg L$, whose generatrix is parallel to the $z$-axis. Eqs. (\ref{eq.betas}) are solved numerically for an incident plane wave $\mathbf {E}_0(\mathbf {r}) = \mathbf {u}_0 E_0 \exp(i k z - i\omega t)$. The transmitted electric field $\mathbf {E}(\mathbf {r})$ at an arbitrary point behind the atomic ensemble can be calculated as follows\footnote{Note that we exclude the contributions of atoms near the lateral edges of the cylinder to avoid finite-size effects.}:
\begin{eqnarray}\label{eq.field}
\begin{aligned}
\mathbf{E}_\mu(\mathbf{r}) &= \mathbf{E}_{0\mu}(\mathbf{r}) - 4\pi k^2 \sum\limits_{i, m, \nu}\mathbf{d}_{g_i e_{im}}^{\nu}G_{\mu\nu}\left(\mathbf{r}-\mathbf{r}_i \right)
\bm{\beta}_{e_{im}}.
\end{aligned}
\end{eqnarray}

Numerically, we assume that the photodetector plane is perpendicular to the $z$-axis and is located at a distance of 15-30 $k^{-1}$, depending on the ensemble radius.
Due to the finite radius $R$, the electric field strength at different points in this plane is different. However, we have checked that at distances less than $R/2$ from the $z$-axis, the field strength averaged over random spatial atomic configurations is constant to a very good approximation. For this reason, to estimate the coherent transmission, we use the following expression, averaged over the atomic ensemble and the photodetector area (radius $R_\mathrm{d}$):
\begin{eqnarray}
t = \frac{1}{\pi R_\mathrm{d}^2 E_0} \int\limits_{\pi R_\mathrm{d}^2} d^2 \mathbf{r} \langle \mathbf{u}_0 \mathbf{E}(\mathbf{r}) \rangle \; .
\end{eqnarray}
The radius of the averaging region
is chosen as $R_\mathrm{d}=R/2$.

To calculate the amplitude reflection coefficient, we use the same method, but the source radiation is not taken into account when calculating the corresponding field strength (Eq. \ref{eq.field}).

\section{Comparison between the two models}\label{sec.comparison}

We first compare the results of the two models at low density, and then we study what happens at larger density.

\subsection{Single layer at low density}\label{sec.single_layer}

We consider a single layer with a cylindrical shape of radius $R = 600 k^{-1}$. The thickness of the layer is $L$ and the atomic density $\rho$. The layer is driven by a plane wave perpendicular to it and we compute the transmission $t$ and reflection $r$ coefficients from the CD model as described above. Then we use the transfer matrix of a single layer to relate the obtained coefficients to the refractive index of the atomic sample via Eqs. (\ref{eq.t}) and (\ref{eq.r}). The two equations each provide two independent equations for the real and imaginary parts of $n$, which are solved numerically.

\begin{figure}[t]
\centering
\includegraphics{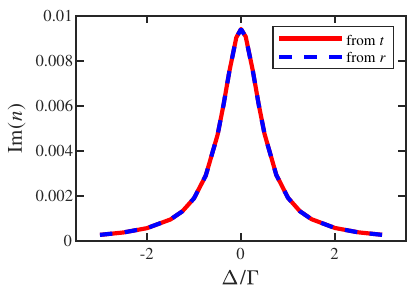}
\caption{Imaginary part of the refractive index in the medium computed from the reflection and transmission by a thin layer of atoms ($kL=0.25$). The reflection and transmission are computed from the CD model.}
\label{fig.low_thin}
\end{figure}

Following this procedure, we first consider a thin layer with $kL = 0.25$ and $\rho = 0.001 k^3$. We plot the imaginary part of the refractive index obtained from either $t$ or $r$ in Fig. \ref{fig.low_thin}. The two results perfectly coincide and they also match the textbook formula
\begin{equation}
\mathrm{Im}(n) = \frac{\rho}{2} \mathrm{Im}(\alpha) \, ,
\end{equation}
where $\alpha$ is the polarizability given by Eq.\,(\ref{eq.alpha}).

We have also checked that it works equally well with the real part. This first benchmark shows that at low density, the CD model is well able to simulate the textbook refractive index, as measured from its macroscopic effect, namely the transmission and reflection .

This result also emphasizes the fact that the refractive index, which is a macroscopic quantity, can be used to describe light propagation in a very thin slice of dilute disordered atoms, provided that we only consider the coherent transmission and reflection averaged over the disorder configurations. This is somewhat counterintuitive as the refractive index is often thought of as being a property of continuous media, which is far from being the case here (see also the discussion in \cite{Guerin:2017a}).



\begin{figure}[t]
\centering
\includegraphics{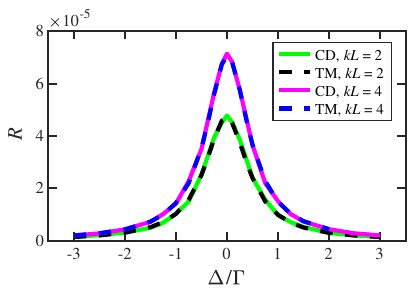}
\caption{Reflection coefficient by a thick layer of atoms, computed from the CD model (solid lines) and from the TM model (dashed lines). The layer thickness is $k L=2$ (green and black) and $k L=4$ (pink and blue).}
\label{fig.low_thick}
\end{figure}

Then we perform a similar comparison with thicker layers. We now plot the reflection coefficient $R=|r|^2$, computed either from the CD model, or from the TM model using the textbook refractive index, with single layers of thickness $kL=2$ and $kL=4$ (Fig.\,\ref{fig.low_thick}). 
The difference between the two models does not exceed the errors of the numerical calculation. We have checked that the agreement between the two models is also perfect with the transmission coefficient $T=|t|^2$. We conclude that the TM method works very well for dilute media. The optical properties of a thick layer can be deduced from the properties of a thin one. 

\subsection{Single layer at large density}\label{sec.large_density}

Let us now investigate the same problem at higher density.

\begin{figure}[t]
\centering
\includegraphics{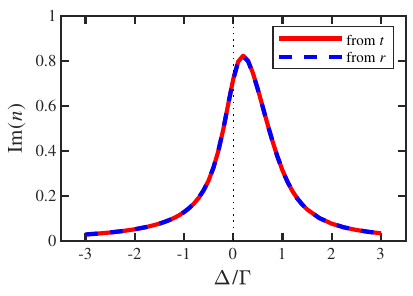}
\caption{Same as in fig.\,\ref{fig.low_thin} but at higher density, $\rho = 0.1 k^3$ instead of $\rho = 0.001 k^3$.}
\label{fig.high_density_n}
\includegraphics{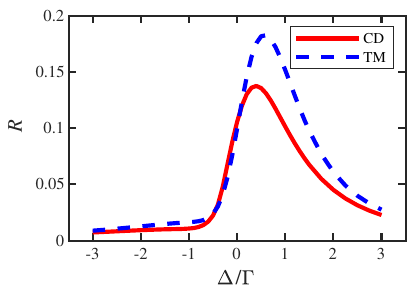}
\caption{Reflection coefficient of the thick ($kL=2$) and dense ($\rho = 0.1 k^3$) atomic layer, computed from two methods. The solid line shows the microscopic CD simulation and the dashed line shows the result of the TM model that uses the refractive index determined from a thin layer.}
\label{fig.high_density_comparison}
\end{figure}

As previously, we compute the complex refractive index based on the microscopic calculation of the transmission and reflection using a thin layer ($kL=0.25$), but now with $\rho = 0.1 k^3$. The result for the imaginary part is shown in Fig.\,\ref{fig.high_density_n}. The results from the transmission and reflection still match, but we observe a shift from the atomic resonance, showing that the textbook Eq.\,(\ref{eq.n}) is not valid any more.

Can we use this effective refractive index to compute the response of a thicker layer? We inject this refractive index, determined from the CD simulation of a thin layer, into the TM model of a thick layer ($kL=2$), and compare with the CD simulation of the thick layer. The comparison is shown in Fig.\,\ref{fig.high_density_comparison}. The discrepancy between the two models appears very clearly.

\begin{figure}
\centering
\includegraphics{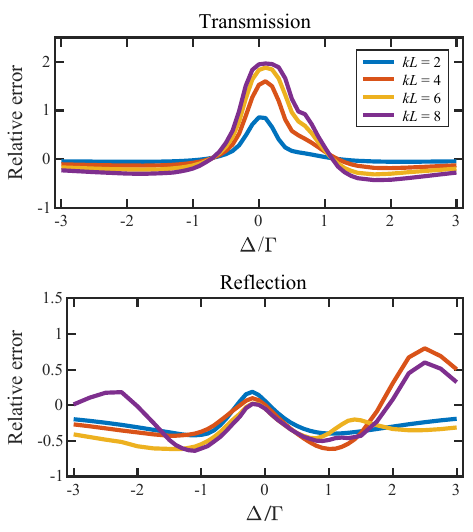}
\caption{Relative error between the CD and the TM models, for the transmission (top) and the reflection (bottom), as a function of the layer thickness. The density is $\rho=0.1k^3$.}
\label{fig.high_density_error}
\end{figure}

In Fig.\,\ref{fig.high_density_error}, we study this discrepancy in more detail, for the transmission and reflection, as a function of the layer thickness. The density is fixed at $\rho = 0.1k^3$. We plot the relative error, i.e. the ratio of the difference in the results to their mean. The relative error increases with the thickness, in particular for the transmission, and can exceed 100\%.


We have repeated the same analysis for a twice larger density, $\rho=0.2k^3$, and the relative error is even larger.

These results show  that we cannot use the results obtained for thin layers for subsequent simulations of thick layers. The layering method leads to errors in both the transmission and reflection. The physical explanation is that atoms from different layers influence each other: the optical properties of a given layer change in the presence of a neighboring one. Thus, we are dealing with a collective response.

\begin{figure}
\centering
\includegraphics{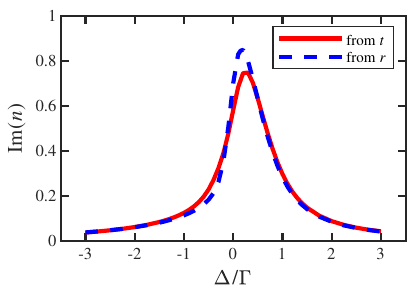}
\caption{Determination of the imaginary part of the refractive index by using the reflection (solid line) or the transmission (dashed line) from CD simulations of a thick layer ($kL=1$). The density is $\rho=0.1k^3$.}
\label{fig.high_thick}
\end{figure}

One could argue that the refractive index determined from a thin layer is not the true refractive index, which is bulk property and should thus be determined from the simulation of a thick sample. Trying to do so raises new problems. Indeed, the CD model does not allow to determine the refractive index directly, but only through its macroscopic consequences, for example the transmission through or the reflection by the sample, as we did previously with thin samples \cite{note:index}. However, with thick and dense samples, one obtains different results if one computes the refractive index from the transmission or from the reflection. This is what is illustrated in Fig.\,\ref{fig.high_thick} with $kL=1$.



\begin{figure}
\centering
\includegraphics{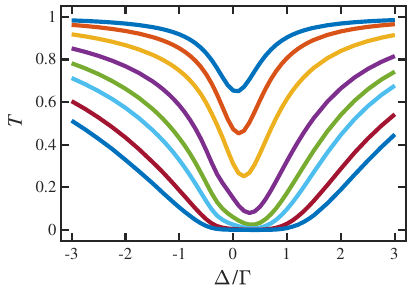}
\caption{Transmission spectra for a fixed density $\rho=0.1k^3$ and varying layer thickness: $kL = 0.25, 0.5, 1, 2, 3, 4, 6, 8$ from top to bottom.}
\label{fig.size_effect}
\end{figure}

Not only the result for the refractive index depends whether we use the transmission or the reflection, but it also depends on the thickness of the sample. This size effect is illustrated in Fig.\,\ref{fig.size_effect}, where we observe a shift of the transmission spectrum as the layer thickness increases. This is reminiscent of some early experiments on this topic, for instance Refs. \cite{Chomaz:2012,Pellegrino:2014}. This finite-size effect can be explained by the fact that the response of each atom is influenced by its neighbors. Therefore, the medium becomes inhomogeneous, as the atoms on the boundaries have a different environment compared to the atoms near the center.



%

\begin{figure}
\centering
\includegraphics{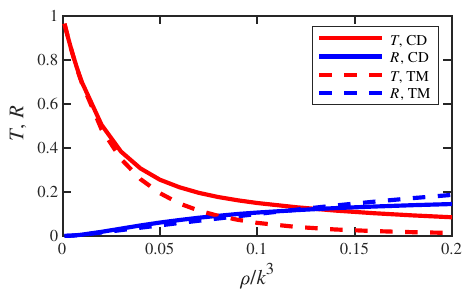}
\caption{Transmission (red) and reflection (blue) coefficients as a function of the density for a fixed detuning $\Delta=0$ and a fixed thickness $kL=0.2$. Solid lines are the CD simulations, and the dashed lines are the TM model applied to the thick layer with the refractive index determined from a thin layer.}
\label{fig.versus_density}
\end{figure}

We conclude that the layer separation method is not applicable to dense atomic ensembles.
For a given thickness and detuning, one can study how the transmission and reflection evolve with the density. We plot this in Fig.\,\ref{fig.versus_density}, again comparing the two methods: the results of the CD simulation for the transmission and reflection coefficients are compared to the one obtained from the TM model in which we used the susceptibility determined from a thin layer as in Fig.\,\ref{fig.low_thin}. The error of the TM approach becomes very noticeable already for density $\rho \sim 0.04 k^3$, with a relative error of about 20\% on the reflection and transmission.


\section{Many layers: case of an atomic Bragg mirror}\label{sec.Bragg}

We now turn to the more complicated case of several atomic layers separated by vacuum, in such a way that it builds a periodic pattern of periodicity close to $\lambda/2$. This is a case of experimental interest, as the constructive interference in the backward direction can produce a strong reflectivity. Such an atomic Bragg mirror has already been demonstrated \cite{Schilke:2011, Schilke:2012b, Asselie:2025} and the TM model has been used to simulate these experiments. Since the separation between layers is smaller than $\lambda/2$, it is interesting to check the validity of this approach, which will depend on the density in each layer.

\begin{figure}
\centering
\includegraphics{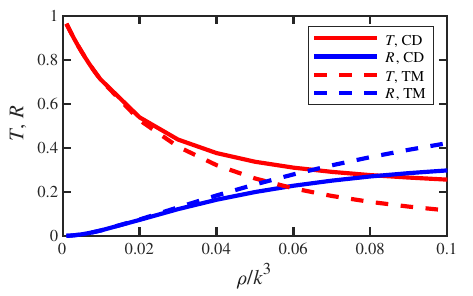}
\caption{Same as in Fig. \ref{fig.versus_density} but in the case of an alternating pattern of thin atomic layer and vacuum in the Bragg configuration.}
\label{fig.Bragg}
\end{figure}

We consider ten layers of thickness $k L=0.2$, separated by a gap of length $\Delta L$ such that $k \Delta L=\pi-0.2$, i.e., the periodicity of the lattice is exactly $\lambda/2$, the Bragg condition. The density of the layers is changed from $0.001k^3$ to $0.1k^3$. Due to the numerical limitations of the CD model, the radius of the layers have to be adapted to the chosen density: for $\rho<0.005k^3$, $k R=500$, for $0.005k^3<\rho<0.009k^3$, $k R=300$, and for $0.01k^3 <\rho < 0.1 k^3$, $k R=100$. The calculations, as before, are carried out in two ways. The first is microscopic with the CD model. The second is based on the TM in which we use the values of the complex susceptibility from the microscopic calculation of the reflection and transmission coefficients of one thin layer. The comparison of the results is shown in Fig.\,\ref{fig.Bragg}. Here, the reflectivity of the Bragg structure is limited by the relatively small number of layers and their small thickness, which are ultimately limited by the number of atoms one can simulate in the CD model. More involved simulations of such a system have been presented in \cite{Samoylova:2014}.


The results are very close to the one of a thick layer (Fig.\,\ref{fig.versus_density}): above $\rho=0.05k^3$, the error of the TM model becomes very significant ($\sim 20\%$). However, it works very well at low density, which validates its use in the experiment \cite{Schilke:2011}, where the peak (resp. mean) density in each layer (having a Gaussian density distribution) was on the order of $\sim 2.4\times 10^{12}$\,cm$^{-3}$ (resp. $\sim 0.7\times 10^{12}$\,cm$^{-3}$), which corresponds to $\sim 0.005 k^3$ (resp. $\sim 0.0016 k^3$) for rubidium with $\lambda = 780$\,nm. In the more recent experiment \cite{Asselie:2025}, the densities were twice larger, which is still well in the applicability range of the TM approach.

\section{Conclusion}

To summarize, we have studied the validity range of the transfer matrix approach for stratified atomic media, 
showing that it works well at low density, including for thin layers, whereas it fails at high density. First, the textbook formulas for the refractive index cannot be used any more, which makes this approach much less useful. Second, and worse, one cannot inject the refractive index obtained from the exact simulation of a thin layer in the transfer matrix model to compute the collective behavior of a thick layer or of many layers. Typically, we find that the transfer-matrix approach leads to a relative error on the order of 20\% on the reflection and transmission coefficients when the atomic density of the individual layers reaches about $0.05 k^3$ (corresponding to $\rho=2.6\times10^{12}$\,cm$^{-3}$ for rubidium atoms).

These results are important to investigate the photonic properties of periodic atomic systems at densities that are significantly higher then those of previous realizations, for example based on ultra-cold atoms (Bose-Einstein condensates or close) \cite{Zhu:2022, Wang:2023}.

\begin{backmatter}
\bmsection{Funding} ISM asknowledges the financial support of the Ioffe Institute within the framework of the baseline project FFUG-2024-0039.
WG acknowledges the support of the French National Research Agency within the framework of the projects 1DOrder (ANR-22-CE47-0011), QUTISYM (ANR-23-PETQ-0002) and E-CANNON (ANR-25-CE47-0361). 


\bmsection{Disclosures} The authors declare no conflicts of interest.

\bmsection{Data availability} Data underlying the results presented in this paper are not publicly available at this time but may be obtained from the authors upon reasonable request.

\end{backmatter}


\end{document}